\documentclass[11pt]{report}

\usepackage{authblk}
\usepackage{bera}
\usepackage{libertine}
\usepackage[libertine]{newtxmath}
\usepackage[utf8]{inputenc}
\usepackage[hmargin=2.5cm,vmargin=2.5cm]{geometry}
\usepackage{listings}
\usepackage{booktabs}
\usepackage{tikz}
\usepackage[justification=centering]{caption}
\usepackage{xcolor,soul}
\usepackage{pgfplots}
\usepackage{pgfmath}
\usepackage{multirow}
\usepackage{mathtools}
\usetikzlibrary{calc}
\usetikzlibrary{shapes,arrows}
\usepackage{filecontents}
\usepackage{amsmath,bm}
\usepackage{amsfonts}
\usepackage{amssymb}
\usepackage{graphicx}
\usepackage{multirow}
\usepackage{wrapfig}
\usepackage{epsfig}
\usepackage{framed}
\usepackage{textcomp}
\usepackage{color}
\usepackage{booktabs}
\usepackage{pdflscape}
\usepackage[usenames,dvipsnames]{pstricks}
\usepackage{cite}
\usepackage{algorithm}
\usepackage[noend]{algpseudocode}
\usepackage{acronym}
  \newacro{cnn}[CNN]{Convolutional Neural Network}
  \newacro{pe}[PE]{Processing Element}
  \newacro{simd}[SIMD]{Single Instruction on Multiple Data}
  \newacro{dsp}[DSP]{Digital Signal Processing}
  \newacro{hdl}[HDL]{Hardware Description Language}
  \newacro{le}[LE]{Logic Elements}
  \newacro{fpga}[FPGA]{Field-Programmable Gate Array}
  \newacro{fifo}[FIFO]{First-In First-Out}
  \newacro{gpu}[GPU]{Graphics Processing Unit}
  \newacro{haddoc}[HADDOC]{Hardware Automated Dataflow Description Of CNNs}
  \newacro{hls}[HLS]{High-Level Synthesis}
  \newacro{moc}[MoC]{Model of Computation}
  \newacroplural{moc}[MoC]{Models of Computation}
  \newacro{ocr}[OCR]{Optical Character Recognition}
  \newacro{qos}[QoS]{Quality of service}
  \newacroplural{qos}[QoS]{Qualities of service}
  \newacro{tpr}[TPR]{True Positive Rate}
  \newacro{mac}[MAC]{Multiply Accumulate}
  \newacro{le}[LE]{Logic Element}
  \newacro{fc}[FC]{Fully Connected}
  \newacro{simd}[SIMD]{Single Instruction on Multiple Data}
  \newacro{vhdl}[VHDL]{VHSIC Hardware Description Language}
  \newacro{lut}[LUT]{Look-Up Table}
  \newacro{nef}[NEF]{Neighborhood Extraction Factorization}  
  \newacro{ne}[NE]{Neighborhood Extraction} 
  \newacro{hdl}[HDL]{Hardware Description Language}
  \newacro{rtl}[RTL]{Register Transfer Level}
  \newacro{ip}[IP]{Intellectual Property}
  \newacro{dhm}[DHM]{Direct Hardware Mapping}
  \newacro{dag}[DAG]{Direct Acyclic Graph}
  \newacro{sdfg}[SDFG]{Synchronous DataFlow Graph}
  \newacro{dpn}[DPN]{modeled as dataflow process network}
  \newacroplural{ip}[IP]{Intellectual Properties}
\usepackage{xcolor}

\colorlet{punct}{red!60!black}
\definecolor{background}{rgb}{0.95,0.95,0.95}
\definecolor{delim}{RGB}{20,105,176}
\definecolor{codegreen}{rgb}{0,0.4,0}
\definecolor{codegray}{rgb}{0.5,0.5,0.5}
\definecolor{codepurple}{rgb}{0.58,0,0.82}

\lstdefinelanguage{vhdl}{
    morekeywords=[1]{
    library,use,all,entity,is,port,generic,map,in,out,end,architecture,of,
    begin,and,or,Not,downto,ALL
   },
   morekeywords=[2]{
     STD_LOGIC_VECTOR,STD_LOGIC,IEEE,STD_LOGIC_1164,
     NUMERIC_STD,STD_LOGIC_ARITH,STD_LOGIC_UNSIGNED,std_logic_vector,
     std_logic
   },
   morecomment=[l]--
}
\lstdefinelanguage{my_vhdl}{
    language     = vhdl,
    basicstyle=\small\fontfamily{fvm}\selectfont,
    showstringspaces=false,
    breaklines=true,
    backgroundcolor=\color{background},
    keywordstyle = [1]\color{codegreen},
    keywordstyle = [2]\color{blue},
    commentstyle=\color{codegreen},
    breaklines=true
    }
\lstdefinelanguage{json}{
    language=C,
    basicstyle=\small\fontfamily{fvm}\selectfont,
    showstringspaces=false,
    breaklines=true,
    backgroundcolor=\color{background},
    commentstyle=\color{codegreen},
    keywordstyle=\color{magenta},
    numberstyle=\tiny\color{delim},
    stringstyle=\color{codegreen}
}

\title{\LARGE{{Hardware Automated Dataflow Deployment of CNNs}}\\ \bigskip
       \Large {Technical Report Haddoc/2016-06TR03}} \bigskip \bigskip \bigskip
\author[1]{K.Abdelouahab}
\author[1,2]{M.Pelcat}
\author[1]{J.Serot}
\author[3]{C.Bourrasset} 
\author[4]{J.C.Quinton}
\author[1]{F.Berry}

\affil[1]{Institut Pascal,Clermont Ferrand, France}
\affil[2]{IETR, INSA Rennes, France}
\affil[3]{CEPP Atos/Bull, Montpellier, France}
\affil[4]{Laboratoire Jean Kuntzmann, Grenoble, France}
\date{June 2017}

\begin{document}

\maketitle
\abstract{Deep \acp{cnn} are the state of the art systems for image classification and scene understating. However, such techniques are computationally intensive and involve highly regular parallel computation. \acp{cnn} can thus benefit from a significant acceleration in execution time when running on fine grain programmable logic devices. As a consequence, several studies have proposed FPGA-based accelerators for \acp{cnn}. However, because of the huge amount of the required hardware resources, none of these studies directly was based on a \emph{direct} mapping of the \ac{cnn} computing elements onto the FPGA physical resources. In this work, we demonstrate the feasibility of this so-called \emph{direct hardware mapping} approach and discuss several associated implementation issues. As a proof of concept, we introduce the \textsc{haddoc2} open source tool, that is able to automatically transform a \ac{cnn} description into a platform independent hardware description for FPGA implementation.
}
\newpage
\section{Introduction}
Convolutional Neural Networks (CNNs) \cite{lecun2015deep} have become a {\em{de-facto}} standard that increased the robustness and accuracy of machine vision systems. It is possible nowadays to build high performance image classification systems by deploying large-scale, pre-trained \acp{cnn} models. However, this accuracy comes at the price of a high computational cost as state of the art \acp{cnn} may require up to 38 GOP to classify a single frame \cite{canziani2016}. As a result, implementing \acp{cnn} with real-time constraints is challenging task. A possible way to address this challenge is to take advantage of the massive fine grain parallelism offered by FPGA devices to embody the large amount of intrinsic parallelism exhibited by \ac{cnn}-based algorithms. In this case, the problem boils down to find an adequate and efficient mapping between the computation model of the latter and the execution model supported by the former. Based on our previous experience in the implementation of real-time vision applications on FPGA-based  platforms~\cite{CAPH}, we advocate the use of a \emph{stream-based dataflow} model to solve this mapping problem. In this approach, a \ac{cnn}-based algorithm is described as  graph of dataflow actors exchanging data through unidirectional channels and this graph is statically and physically mapped onto the target FPGA using a library of pre-defined computing elements to implement actors. 

In the sequel,  we demonstrate the feasibility of this so-called \emph{\ac{dhm}} approach for implementing realistic CNN-based applications onto  \acp{fpga}. Moreover, we introduce \textsc{haddoc2}, a software framework providing a fully automated implementation path for \acp{cnn} onto \acp{fpga} using the \ac{dhm} approach. The \textsc{haddoc2} tool is compatible with the widely used Caffe deep learning framework~\cite{Jia2014} and generates platform independent synthetizable VHDL code. In other words, we introduce in this work a tool that automatically maps a Caffe pre-trained model onto an \ac{fpga} device.
\section{CNNs : Computations and parallelism sources}
\begin{figure}[!h]
	\centering
	\includegraphics[width=0.75\textwidth]{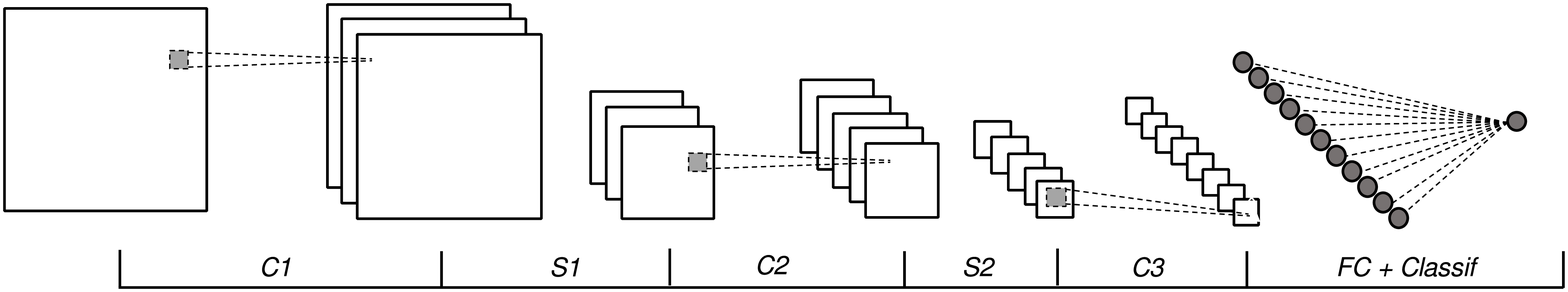}
	\caption{An example of a \ac{cnn} topology with 3 convolutional layers (C1,C2,C3)\\ two subsampling layers and one fully connected stage (FC).}
    \label{cnn_topo}
\end{figure}

\acp{cnn} are a category of feed forward artificial neural networks that are bio-inspired by the visual cortex of the brain. The huge improvement of \ac{cnn}-based algorithms was made possible by two factors: On one hand, the availability of massive-sized annotated image data-sets \cite{imageNet09} allowed to train robust large scale feature extractors and accurate classifiers.
On the other hand, the growth of high performance processors and, especially \acp{gpu}, provided the computational power required to train deeper and more complex neural networks \cite{cudnn14}.  A typical \ac{cnn} structure, as shown in figure \ref{cnn_topo}, will perform a succession of convolutions interspersed with sub-sampling layers. The last stages include typically two or three fully connected neural network for classification tasks. The depth (number of layers) of a \ac{cnn} ensures better accuracy and less over-fitting. As a result, depth of neural networks tend to increase (8 to 19 layers to VGG \cite{vgg14}). 

\subsection{Convolution layers}
\label{overview:conv}
Convolutional layers are the most computationally intensive and are responsible -- in a typical implementation -- for more than 90\% of the \ac{cnn} execution time~\cite{cong14}. Each layer $(l)$ extracts $N$ feature maps from $C$ input channels by performing $N$ convolutions of size $K \times K$ on each input. This filtering is followed by the application of a non-linear activation function $act$ and a bias term $b_n$ to each set of features. As shown in equation~\ref{convLayerProc}, $N \times C$ convolutions are required to process a given layer. 

\begin{align}
    \label{convLayerProc}
    \forall l =1:L& \nonumber \text{\textit{ (Number of $conv$ layers)}}\\
    \forall n =1&:N \nonumber \text{\textit{  (Number of output feature maps)}}\\
    \forall i =&1:Ix \nonumber \text{\textit{  (Feature map rows)}}\\
    \forall j &=1:Iy \nonumber \text{\textit{  (Feature map columns)}}\\          
    & f^{(l)}[n,i,j] = b^{(l)}[n] + \sum_{c=1}^{C} \sum_{p=1}^{K}  \sum_{q=1}^{K} \Phi^{(l)}[c,i+p,j+q] . w^{(l)}[n,c,p,q]
\end{align}

% \begin{align}
%     \label{convLayerProc}
%     \forall l = &1:L   \nonumber \text{\textit{ (Number of $conv$ layers)}}\\
%     \forall n &=1:N^{(l)} \nonumber \text{\textit{  (Number of output feature maps)}}\\
%     & \bm{f_n}^{(l)} = \mbox{act} \left[ \bm{b_n}^{(l)} + \sum_{c=1}^{C^{(l)}} conv(\bm{\phi_c^{(l)}}, \bm{w_{nc}^{(l)}}) \right]
% \end{align}
where
\begin{itemize}
	\item $\bm{{f}^{(l)}}$ is a tensor of output feature maps of layer  $(l)$
	\item $\bm{b^{(l)}[n]}$ is the bias term applied to applied to feature $n$
	\item ${\bm{\Phi^{(l)}}}$ is a tensor of input feature maps of layer  $(l)$
	\item ${\bm{w^{(l)}}}$ is tensor of pre-learned filters
\end{itemize}

% The $conv$ operator is defined in equation \ref{convOperator} and is responsible of the 2D filtering and thus the feature extraction in convolutional layers.   
% \begin{align}
% 	\label{convOperator}
% 	\forall i = 1 :& K   \nonumber \text{\textit{ (input rows)}}\\
% 	\forall j = 1&:K   \nonumber \text{\textit{ (input columns)}}\\
% %    \forall p = &1:K   \nonumber \text{\textit{ (kernel rows)}}\\
% %    \forall q &=1:K    \nonumber \text{\textit{ (kernel columns)}}\\
%   	&conv(\bm{\phi_c^{(l)}}, \bm{w_{nc}^{(l)}}) = \sum_{p=1}^{K^{(l)}}  \sum_{q=1}^{K^{(l)}}  \bm{\phi_c^{(l)}}[i+p,j+q] . {w^{(l)}_{nc}}[p,q]
% \end{align}

As already pointed out in \cite{mot16}, the computations described in equations \ref{convLayerProc}  exhibit a large amount of potential parallelism:\\ 

\begin{itemize}
\item\textbf{{Inter Layer parallelism:}} \acp{cnn} have a feed-forward hierarchical structure consisting of a succession of data-dependent layers. Layers can therefore that can be executed in a \emph{pipelined} fashion where the execution of layer $(l)$ can start before the execution of layer $(l-1)$ ends.

\item\textbf{{Inter neuron parallelism:}} Each neuron of a layer is independent when processing features. Thereby, a full data-parallelism can be exploited when computing concurrently each of the $N^{(l)}$ element of equation \ref{convLayerProc}

\item\textbf{{Inter convolution parallelism:}} All of the convolutions performed by a single neuron can also be evaluated simultaneously by computing concurrently the $C^{(l)}$ convolutions of equation~\ref{convLayerProc}.

\item\textbf{{Intra convolution parallelism:}} 2D image convolution can be implemented in a pipelined fashion~\cite{shoup94} allowing the $K \times K$ multiplications to be computed concurrently in equation \ref{convLayerProc}
    
\end{itemize}

\subsection{Subsampling layers}
A common operation when conceiving \acp{cnn} is to periodically insert subsampling (or pooling) layers in-between successive convolutional layers. These downsample the inputs by selecting the \emph{average}, or, more commonly, the \emph{maximum} of a given neighborhood of each pixel as described in equation \ref{poolProc}

\begin{align}
	\label{poolProc}
    \forall l =1:L& \nonumber \text{\textit{ (Number of $pool$ layers)}}\\
    \forall n =1&:N \nonumber \text{\textit{  (Number of output feature maps)}}\\
    \forall i =&1:Ix \nonumber \text{\textit{  (feature map rows)}}\\
    \forall j &=1:Iy \nonumber \text{\textit{  (feature map columns)}}\\ 
    & f^{(l)}[n,i,j] = \max_{p,q \in [1:K]}{\left(\Phi^{(l)}[n,i+p,j+q]\right)}
\end{align}

Pooling layers reduce the amount of parameters required to process the next stages of the network, which controls overfitting in one hand and decrease the computation load on the other.

\subsection{Fully connected layers}
A \ac{fc} neural network --with usually 3 or 4 hidden layers-- terminates \acp{cnn} and acts as a classifier. In this case, no parameters are shared across the feature-maps (feature maps and learned parameters have the same dimension). In this case, \ac{fc} layer activations are computed with the inner product operation followed by a bias offset as detailed in equation \ref{fcLayerProc}, where $<,>$ denotes the the inner product operator.

\begin{align}
    \label{fcLayerProc}
    \forall l =1:L& \nonumber \text{\textit{ (Number of $FC$ layers)}}\\
    \forall n =1&:N \nonumber \text{\textit{  (Number of output feature maps)}}\\
    &\bm{f^{(l)}[n]} = \mbox{act} \left[ {b^{(l)}[n]} + \sum_{c=1}^{C^{(l)}} <\bm{\phi^{(l)}[c]}, \bm{w^{(l)}[n,c]}> \right]
\end{align}

\section{Direct Hardware Mapping of CNN entities}
\subsection{Dataflow processing of CNNs}

The foundations of dataflow \acp{moc} were formalized by \cite{den74} in order to create an architecture where multiple fragments of instructions can process simultaneously a stream of data. Programs respecting dataflow semantics are described as a \emph{network} (graph) of fundamental processing units commonly called \emph{actors} and communicating abstract data messages called \emph{tokens} on unidirectional \ac{fifo} channels. 

In terms of architecture-application matching, the \ac{cnn}'s layout fits naturally with a stream-based model of computation. All of the operations involved in feed forward propagation of a \ac{cnn} --described in the latter section-- can be executed following the stream-based dataflow \ac{moc}. In fact, \ac{cnn}-based algorithms can be modeled as \acp{dpn} where nodes correspond to processing actors and edges correspond to communication channels. Each actor follows a purely data-driven execution model where execution \emph{(firing)} is triggered only by the availability of input operands. 

The \ac{dhm} approach consists of \emph{physically} mapping entirely graph of actors onto the target device. Each actor becomes a computing unit with its specific instance on the \ac{fpga} and each edge is mapped to a signal.

\subsection{DHM of Convolution layers}
\begin{figure}[!h]
    \hfill
    \begin{minipage}[t]{0.15\linewidth}
        %\flushleft
        \begin{tikzpicture}[thick,scale=0.65, every node/.style={scale=0.75}]

	\tikzset{dashedBox/.style={draw,rectangle,dashed,rounded corners=2pt,minimum width=1.2cm,minimum height=5cm}};
    \tikzset{prod/.style={draw,circle,minimum size=0.5cm}};
    \tikzset{sumB/.style={draw,circle,minimum size=1cm}};
    \tikzset{io/.style={}};
    
    \node [io] (p0) at (0.5,-1.5)  {$\phi_0$};
    \node [io] (p1) at (0.5,-2.0) 	{$\phi_1$};
    \node [io] (pd) at (0.5,-2.5) 	{$\vdots$};
    \node [io] (pk) at (0.5,-3.0) 	{$\phi_C$};

    \node [dashedBox](l) at (2,-2)    {};
    
    \node [prod] (prod0) at (2,0.0)   {$\eta_0$};
    \node [prod] (prod1) at (2,-1.2)  {$\eta_1$};
    \node [io]   (prodd) at (2,-2.4)  {$\vdots$};
    \node [prod] (prodk) at (2,-3.6)  {$\eta_N$};
       
    \node [io]    (f0) 	 at (3.5, 0.0)    	{$f_0$};
    \node [io]    (f1) 	 at (3.5,-1.2)  	{$f_1$};
    \node [io]    (fk) 	 at (3.5,-3.6)  	{$f_N$};
    %\node []      (nimp) at (2.0,-4.3)      {\textbf{$C_1$}};

    \draw[->,>=latex] (p0)--(l.west|-p0);
    \draw[->,>=latex] (p1)--(l.west|-p1);
    \draw[->,>=latex] (pk)--(l.west|-pk);
    \draw[->,>=latex] (l.east|-f0)--(f0);
    \draw[->,>=latex] (l.east|-f1)--(f1);
    \draw[->,>=latex] (l.east|-fk)--(fk);

	\node (title) at (2,-5.5) {\textit{(a)}};     
\end{tikzpicture}
    \end{minipage}
    \hfill
    \begin{minipage}[t]{0.3\linewidth}
        %\centering
        \begin{tikzpicture}[thick,scale=0.65, every node/.style={scale=0.85}]
    
    \tikzset{conv/.style={draw,rectangle,rounded corners=3pt,minimum size=0.5cm}};
    \tikzset{sum/.style={draw,circle,minimum size=0.7cm}};
    \tikzset{act/.style={draw,rectangle,rounded corners=3pt,minimum size=0.8cm}};
    \tikzset{void/.style={}}
    \tikzset{neuronBox/.style={draw,rectangle,dashed,rounded corners=20pt,minimum width=4.2cm,minimum height=3.5cm}};

    \node [void] (il) at (-1.8,-0.7){$\phi_{C}$};
    \node [void] (ii) at (-1.8,0)   {$\vdots$};
    \node [void] (i1) at (-1.8,0.7) {$\phi_1$};
    \node [void] (i0) at (-1.8,1.4) {$\phi_0$};
    \node [void] (b)  at (1.8,2.5)  {$b_0$};

    \node [conv] (cel) at (0,-0.7){$\mbox{conv}_{0C}$};
    \node [void] (cei) at (0,0)   {$\vdots$};
    \node [conv] (ce1) at (0,0.7) {$\mbox{conv}_{01}$};
    \node [conv] (ce0) at (0,1.4) {$\mbox{conv}_{00}$};
    
    \node [sum]  (s)    at (1.8,0.7) {$\Sigma$};
    \node [act]  (tanh) at (3.4,0.7) {{act}};

    \node [neuronBox] (box)  at (1.55,0.7)  {};
    \node [void]      (name) at (1.8,-1.2)  {{$\bm{\eta_}0$}};
    
    \node [void]      (o0) at (5,0.7)   {$f_0$};
    
    \draw[->,>=latex] (i0)--(ce0);
    \draw[->,>=latex] (i1)--(ce1);
    \draw[->,>=latex] (il)--(cel);
    
    \draw[->,>=latex] (b)--(s);
    \draw[->,>=latex] (ce0)--(s);
    \draw[->,>=latex] (ce1)--(s);
    \draw[->,>=latex] (cel)--(s);
    
    \draw[->,>=latex] (s)--(tanh);
    
    \draw[->,>=latex] (tanh)--(o0);
    
    \node (title) at (1.8,-2.3) {\textit{(b)}};     
    
\end{tikzpicture}
    \end{minipage}
    \hfill
    \begin{minipage}[t]{0.3\linewidth}
        %\flushright
        \begin{tikzpicture}[thick,scale=0.65, every node/.style={scale=0.75}]

    \tikzset{prod/.style={draw,circle,minimum size=0.5cm}};
    \tikzset{sumB/.style={draw,circle,minimum size=1cm}};
    \tikzset{io/.style={}};
    \tikzset{neuronBox/.style={draw,rectangle,dashed,rounded corners=2pt,minimum width=2.6cm,minimum height=4.8cm}};
        
    \node [io] (p0) at (0.5,0.2)  	{p00};
    \node [io] (p1) at (0.5,-1)   {p01};
    \node [io] (pd) at (0.5,-2.5)   {$\vdots$};
    \node [io] (pk) at (0.5,-4)     {pkk};
        
    %\node [io] (w0) at (2.2, 0.8) {w00};
    %\node [io] (w1) at (2.2,-0.7) {w01};
    %\node [io] (wk) at (2.2,-3.2) {wkk};
    
    \node [prod] (prod0) at (2.2,0.2)  {$\times$};
    \node [prod] (prod1) at (2.2,-1) {$\times$};
    \node [io]   (prodd) at (2.2,-2.5) {$\vdots$};
    \node [prod] (prodk) at (2.2,-4) {$\times$};
    
    \node [sumB] (s)     at (3.8,-2) {$\sum$};
    
    \draw[->,>=latex] (p0)--(prod0);
    \draw[->,>=latex] (p1)--(prod1);
    \draw[->,>=latex] (pk)--(prodk);
    %\draw[->,>=latex] (w0)--(prod0);
    %\draw[->,>=latex] (w1)--(prod1);
    %\draw[->,>=latex] (wk)--(prodk);
    
    \draw[->,>=latex] (prod0)--(s);
    \draw[->,>=latex] (prod1)--(s);
    \draw[->,>=latex] (prodk)--(s);
    \node (title) at (2.5,-5.5) {\textit{(c)}};  
    
    \node (conv)  at (3.8,-4) {$\bm{\mbox{conv}_{00}}$};
    \node[neuronBox] (box) at (3.1,-1.8) {}; 
    
\end{tikzpicture}
    \end{minipage}
    \hfill
%\centering
\caption{The 3 levels of \ac{dhm} implementation of \ac{cnn} entities:\\ (a) in convolution layers, (b) in neurons, (c) in convolution engines}
\label{dhm_entities}
\end{figure}
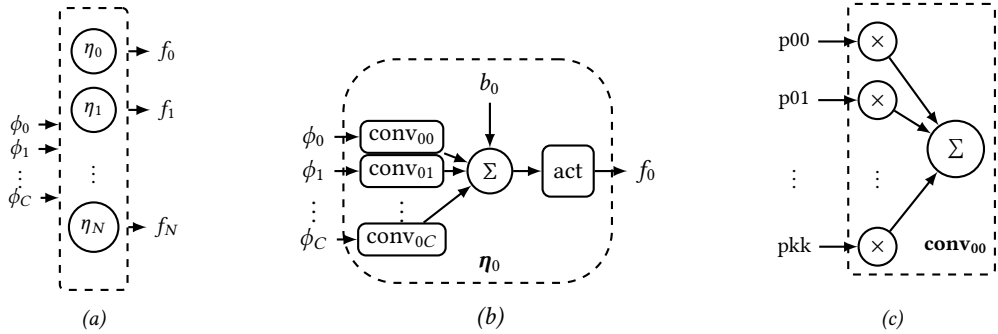

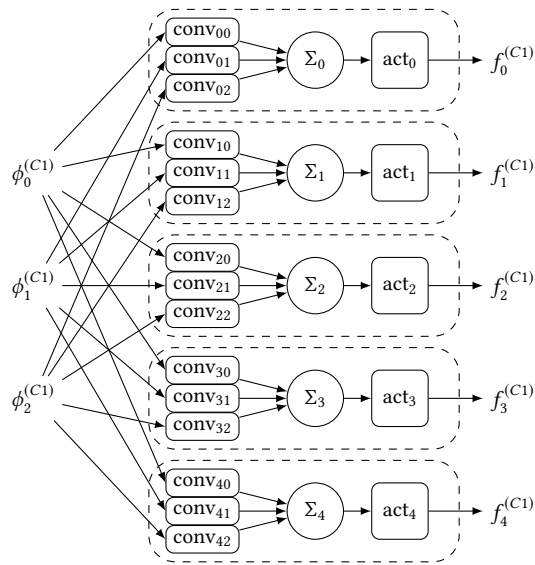
\begin{figure}[!h]
    \centering
    \begin{tikzpicture}[scale=0.75, every node/.style={scale=0.75}]

    \tikzset{neuron/.style={draw,circle,minimum size=1.2cm}};
    \tikzset{conv/.style={draw,rectangle,rounded corners=3pt,minimum size=0.4cm}};
    \tikzset{sum/.style={draw,circle,minimum size=1cm}};
    \tikzset{act/.style={draw,rectangle,rounded corners=3pt,minimum size=1cm}};
    \tikzset{neuronBox/.style={draw,rectangle,dashed,rounded corners=10pt,minimum width=5.5cm,minimum height=1.8cm}};
    \tikzset{void/.style={}}

    \node[void]     (i0)      at(-3,2)   {$\phi^{{(C1)}}_2$};
    \node[void]     (i1)      at(-3,4)   {$\phi^{{(C1)}}_1$};
    \node[void]     (i2)      at(-3,6)   {$\phi^{{(C1)}}_0$};
%    \node[]         (r)       at(2,4)    {\huge{\textcolor{red}{ADD NETLIST-VIEW}}};

    \node [conv] (ce0)  at (0,-0.5)   {{$\mbox{conv}_{42}$}};
    \node [conv] (ce1)  at (0,0)      {{$\mbox{conv}_{41}$}};
    \node [conv] (ce2)  at (0,0.5)    {{$\mbox{conv}_{40}$}};
    \node [conv] (ce3)  at (0,1.5)    {{$\mbox{conv}_{32}$}};
    \node [conv] (ce4)  at (0,2)      {{$\mbox{conv}_{31}$}};
    \node [conv] (ce5)  at (0,2.5)    {{$\mbox{conv}_{30}$}};
    \node [conv] (ce6)  at (0,3.5)    {{$\mbox{conv}_{22}$}};
    \node [conv] (ce7)  at (0,4)      {{$\mbox{conv}_{21}$}};
    \node [conv] (ce8)  at (0,4.5)    {{$\mbox{conv}_{20}$}};
    \node [conv] (ce9)  at (0,5.5)    {{$\mbox{conv}_{12}$}};
    \node [conv] (ce10) at (0,6)      {{$\mbox{conv}_{11}$}};
    \node [conv] (ce11) at (0,6.5)    {{$\mbox{conv}_{10}$}};
	\node [conv] (ce12) at (0,7.5)    {{$\mbox{conv}_{02}$}};
    \node [conv] (ce13) at (0,8)      {{$\mbox{conv}_{01}$}};
    \node [conv] (ce14) at (0,8.5)    {{$\mbox{conv}_{00}$}};
    
    \node[sum]   (sum0)      at(2,0)    {$\Sigma_{4}$};
    \node[sum]   (sum1)      at(2,2)    {$\Sigma_{3}$};
    \node[sum]   (sum2)      at(2,4)    {$\Sigma_{2}$};
    \node[sum]   (sum3)      at(2,6)    {$\Sigma_{1}$};
    \node[sum]   (sum4)      at(2,8)    {$\Sigma_{0}$};
    
    \node[act]   (act0)      at(3.5,0)    {$\mbox{act}_{4}$};
    \node[act]   (act1)      at(3.5,2)    {$\mbox{act}_{3}$};
    \node[act]   (act2)      at(3.5,4)    {$\mbox{act}_{2}$};
    \node[act]   (act3)      at(3.5,6)    {$\mbox{act}_{1}$};
    \node[act]   (act4)      at(3.5,8)    {$\mbox{act}_{0}$};

    \node[void]     (o0)      at(5.5,0)  {$f^{{(C1)}}_4$};
    \node[void]     (o1)      at(5.5,2)  {$f^{{(C1)}}_3$};
    \node[void]     (o2)      at(5.5,4)  {$f^{{(C1)}}_2$};
    \node[void]     (o3)      at(5.5,6)  {$f^{{(C1)}}_1$};
    \node[void]     (o4)      at(5.5,8)  {$f^{{(C1)}}_0$};
    
    \node[neuronBox]   (n0)      at(1.8,0)  {};
    \node[neuronBox]   (n1)      at(1.8,2)  {};
    \node[neuronBox]   (n2)      at(1.8,4)  {};
    \node[neuronBox]   (n3)      at(1.8,6)  {};
    \node[neuronBox]   (n4)      at(1.8,8)  {};
    
    \draw[->,>=latex] (i0)--(ce0.west);
    \draw[->,>=latex] (i0)--(ce3.west);
    \draw[->,>=latex] (i0)--(ce6.west);
    \draw[->,>=latex] (i0)--(ce9.west);
    \draw[->,>=latex] (i0)--(ce12.west);

    \draw[->,>=latex] (i1)--(ce1.west);
    \draw[->,>=latex] (i1)--(ce4.west);
    \draw[->,>=latex] (i1)--(ce7.west);
    \draw[->,>=latex] (i1)--(ce10.west);
    \draw[->,>=latex] (i1)--(ce13.west);

    \draw[->,>=latex] (i2)--(ce2.west);
    \draw[->,>=latex] (i2)--(ce5.west);
    \draw[->,>=latex] (i2)--(ce8.west);
    \draw[->,>=latex] (i2)--(ce11.west);
    \draw[->,>=latex] (i2)--(ce14.west);

    \draw[->,>=latex] (ce0) --(sum0)   ;
    \draw[->,>=latex] (ce3) --(sum1)   ;
    \draw[->,>=latex] (ce6) --(sum2)   ;
    \draw[->,>=latex] (ce9) --(sum3)   ;
    \draw[->,>=latex] (ce12)--(sum4)   ;
    \draw[->,>=latex] (ce1) --(sum0)   ;
    \draw[->,>=latex] (ce4) --(sum1)   ;
    \draw[->,>=latex] (ce7) --(sum2)   ;
    \draw[->,>=latex] (ce10)--(sum3)   ;
    \draw[->,>=latex] (ce13)--(sum4)   ;
    \draw[->,>=latex] (ce2) --(sum0)   ;
    \draw[->,>=latex] (ce5) --(sum1)   ;
    \draw[->,>=latex] (ce8) --(sum2)   ;
    \draw[->,>=latex] (ce11)--(sum3)   ;
    \draw[->,>=latex] (ce14)--(sum4)   ;

    \draw[->,>=latex] (sum0)--(act0);
    \draw[->,>=latex] (sum1)--(act1);
    \draw[->,>=latex] (sum2)--(act2);
    \draw[->,>=latex] (sum3)--(act3);
    \draw[->,>=latex] (sum4)--(act4);

    \draw[->,>=latex] (act0)--(o0);
    \draw[->,>=latex] (act1)--(o1);
    \draw[->,>=latex] (act2)--(o2);
    \draw[->,>=latex] (act3)--(o3);
    \draw[->,>=latex] (act4)--(o4);
    
\end{tikzpicture} 
    \caption{Applying the 3 levels of DHM (fig~\ref{dhm_entities}) to a dummy convolutional layer C1 (N=5, C=3, K=3):\\ \small{15 separate convolution engines (135 Multipliers and 15 adders) plus 5 adders and 5 activation blocks\\ are required to process the layer in a full parallel fashion. (bias omitted)}}
    \label{dhm_layer}
\end{figure}

As stated in section~\ref{overview:conv}, convolutional layers are the most computation intensive tasks in a given network. However, \ac{dhm} approach fully exploits all the parallelism sources of theses layers. All neurons of a layer are mapped on the device to take advantage of intra-neuron parallelism (Fig~\ref{dhm_entities}-a). In neurons, each convolution is mapped separately (Fig~\ref{dhm_entities}-b) and finally, within a convolution engine, each multiplier is instantiated separately (Fig~\ref{dhm_entities}-c). As an example, figure~\ref{dhm_layer} illustrates how a convolution layer C1 ($C=3, N=5, K=3$) extracts 5 features from a 3-channel input pixel flow. In this example, 15 convolution and 5 activation blocks are mapped onto the \ac{fpga} as a result of the layer graph transformation, which corresponds to 135 multiplications, 20 summations and 5 activations.  

%\subsection{DHM of Subsampling layers}
%\subsection{DHM of Fully connected layers}

\section{Optimizing DHM-based CNN accelerators}
Direct Hardware Mapping of \acp{cnn} completely removes the need for an external memory to store intermediate results or parameters. Moreover, thanks to the fully pipelined execution model, the global throughput is only limited by the maximum clock frequency. However, these advantages come at the cost of a high resource consumption since the whole graph has to mapped onto the physical resources of the FPGA. In certain cases, this could limit the complexity of the CNNs that can be handled by the DHM approach. It is crucial, therefore, to ensure that the core operations involved in CNN actors can be translated efficiently in hardware. The most important issues, by far, are those related to on-chip memory requirements on one hand, and the implementation of arithmetic operators on the other.

\subsection{Neighborhood extraction}

The literature provides multiple approach to efficiently accelerate the computation of convolutions. Dataflow-based based accelerators --such in \cite{shoup94}-- are based on a fully pipelined architecture that is able to process one convolution per clock cycle. Such an architecture can be divided into 2 parts: neighborhood extraction (NE) and Multiply-ACCumulation (MAC). 

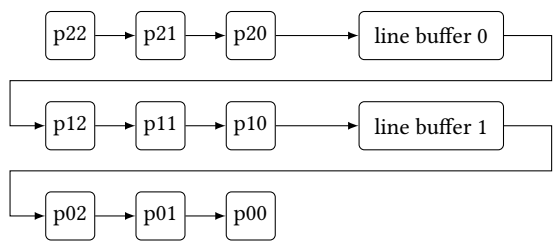
\begin{figure}[!h]
    \centering
    \begin{tikzpicture}[scale=0.8, every node/.style={scale=0.8}]
    \tikzset{latch/.style={draw,rectangle,rounded corners=2pt,minimum size=0.8cm}};
    \tikzset{taps/.style={draw,rectangle,rounded corners=2pt,minimum width=2.4cm,minimum height=0.8cm}};
    \tikzset{void/.style={}}
    
    % \node [rotate=90] (in0) at (-1.5,1.5){in\_flow};
    % \node (i0) at (-1,3.5) {\texttt{in}};
    % \draw[->,>=latex] (i0)|-(p00);
    \node (o0) at (8.5,3.0) {};
    \node (o1) at (8,1.5) {};

    \node [latch] (p20)   at (0.0,0.0)   {p02};
    \node [latch] (p21)   at (1.5,0.0)   {p01};
    \node [latch] (p22)   at (3.0,0.0)   {p00};
    
    \node [latch] (p10)   at (0.0,1.5)   {p12};
    \node [latch] (p11)   at (1.5,1.5)   {p11};
    \node [latch] (p12)   at (3.0,1.5)   {p10};
    \node [taps]  (line1) at (6.0,1.5)   {line buffer 1};
    
    \node [latch] (p00)   at (0.0,3.0)   {p22};
    \node [latch] (p01)   at (1.5,3.0)   {p21};
    \node [latch] (p02)   at (3.0,3.0)   {p20};
    \node [taps]  (line0) at (6.0,3.0)   {line buffer 0};
    
    \draw[->,>=latex] (p20)--(p21);
    \draw[->,>=latex] (p21)--(p22);
    
    \draw[->,>=latex] (p10)--(p11);
    \draw[->,>=latex] (p11)--(p12);
    \draw[->,>=latex] (p12)--(line1);
    \draw[->,>=latex] (line1)-|(8cm,0.75cm)--(-1.0cm,0.75cm)|-(p20.west);
    
    \draw[->,>=latex] (p00)--(p01);
    \draw[->,>=latex] (p01)--(p02);
    \draw[->,>=latex] (p02)--(line0);
    \draw[->,>=latex] (line0)-|(8cm,2.25cm)--(-1.0cm,2.25cm)|-(p10.west);

\end{tikzpicture}
    \caption{Architecture of a 3 $\times$ 3 neighborhood extractor : 2 Buffers with image length\\ size are required to perform a 3 $\times$ 3 convolution on streams of pixels $p_{ij}$}
    \label{ne_archi}
\end{figure}

\ac{ne} relies on buffers to grant a full access to the $K^{(l)} \times K^{(l)}$ neighbors of each pixel (as shown in figure \ref{ne_archi}). Such an architecture is advantageous since it can directly extract the neighborhood of streams of pixels each clock-cycle.

\ac{mac} performs a multiplication of neighborhood pixels with pre-learned kernels then accumulates the result to output feature maps. As long as the access to full neighborhood pixels is guaranteed, each of the multiplications of can be performed in a parallel way using $K^{(l)} \times K^{(l)}$ multipliers (as shown in Fig~\ref{dhm_entities}-c).

In the case of CNNs, Combining \ac{ne} and parallel \ac{mac} strategy fully exploits the intra Kernel parallelism of \acp{cnn} which grants high acceleration to convolutions and, consequently, the feature extraction process. However, mapping a full CNN graph involving millions of convolutions comes down to map millions of memory buffers on the FPGA fabric which increases the power consumption of the system and lowers the maximum frequency (and thus the computation throughput).

\subsection{\ac{nef}}
One way to address the latter issue is to factorize the neighborhood extraction process in order to optimize the memory print of convolutional layers. In this case, it is possible to rely only on on-chip memory buffers to process a hole convolutional layer. 

\begin{figure}[!h]
    \centering
    \begin{tikzpicture}[scale=0.78, every node/.style={scale=0.78}]

    \tikzset{neuron/.style={draw,circle,minimum size=1.2cm}};
    \tikzset{mac/.style={draw,rectangle,rounded corners=3pt,,minimum size=0.4cm,minimum height=0.4cm}};
	\tikzset{ne/.style={draw,rectangle,rounded corners=3pt,minimum size=1.0cm}};
    \tikzset{sum/.style={draw,circle,minimum size=1cm}};
    \tikzset{act/.style={draw,rectangle,rounded corners=3pt,minimum size=1cm}};
    \tikzset{neuronBox/.style={draw,rectangle,dashed,rounded corners=10pt,minimum width=5.5cm,minimum height=1.8cm}};
    \tikzset{void/.style={}}

    \node[void]     (i0)      at(-3.5,0.5)  {$\phi^{(C1)}_2$};
    \node[void]     (i1)      at(-3.5,4)    {$\phi^{(C1)}_1$};
    \node[void]     (i2)      at(-3.5,7.5)  {$\phi^{(C1)}_0$};

	\node [ne]     (ne0)      at(-2,0.5)  {{ne}};
    \node [ne]     (ne1)      at(-2,4)    {{ne}};
    \node [ne]     (ne2)      at(-2,7.5)  {{ne}};

    \node [mac] (mac0)  at (-0.5,-0.5)   {{mac}};
    \node [mac] (mac1)  at (-0.5,0)      {{mac}};
    \node [mac] (mac2)  at (-0.5,0.5)    {{mac}};
    \node [mac] (mac3)  at (-0.5,1)      {{mac}};
    \node [mac] (mac4)  at (-0.5,1.5)    {{mac}};
    \node [mac] (mac5)  at (-0.5,3)      {{mac}};
    \node [mac] (mac6)  at (-0.5,3.5)    {{mac}};
    \node [mac] (mac7)  at (-0.5,4)      {{mac}};
    \node [mac] (mac8)  at (-0.5,4.5)    {{mac}};
    \node [mac] (mac9)  at (-0.5,5)      {{mac}};
    \node [mac] (mac10) at (-0.5,6.5)    {{mac}};
    \node [mac] (mac11) at (-0.5,7)      {{mac}};
	\node [mac] (mac12) at (-0.5,7.5)    {{mac}};
    \node [mac] (mac13) at (-0.5,8)      {{mac}};
    \node [mac] (mac14) at (-0.5,8.5)    {{mac}};
    
    \node[sum]   (sum0)      at(2,0)    {$\sum$};
    \node[sum]   (sum1)      at(2,2)    {$\sum$};
    \node[sum]   (sum2)      at(2,4)    {$\sum$};
    \node[sum]   (sum3)      at(2,6)    {$\sum$};
    \node[sum]   (sum4)      at(2,8)    {$\sum$};
    
    \node[act]   (act0)      at(3.5,0)    {{{act}}};
    \node[act]   (act1)      at(3.5,2)    {{{act}}};
    \node[act]   (act2)      at(3.5,4)    {{{act}}};
    \node[act]   (act3)      at(3.5,6)    {{{act}}};
    \node[act]   (act4)      at(3.5,8)    {{{act}}};

    \node[void]     (o0)      at(5,0)  {$f^{(C1)}_4$};
    \node[void]     (o1)      at(5,2)  {$f^{(C1)}_3$};
    \node[void]     (o2)      at(5,4)  {$f^{(C1)}_2$};
    \node[void]     (o3)      at(5,6)  {$f^{(C1)}_1$};
    \node[void]     (o4)      at(5,8)  {$f^{(C1)}_0$};

    \draw[->,>=latex] (i0)--(ne0);
    \draw[->,>=latex] (i1)--(ne1);
    \draw[->,>=latex] (i2)--(ne2);
	
    \draw[->,>=latex] (ne0)--(mac0.west);
    \draw[->,>=latex] (ne0)--(mac1.west);
    \draw[->,>=latex] (ne0)--(mac2.west);
    \draw[->,>=latex] (ne0)--(mac3.west);
    \draw[->,>=latex] (ne0)--(mac4.west);

    \draw[->,>=latex] (ne1)--(mac5.west);
    \draw[->,>=latex] (ne1)--(mac6.west);
    \draw[->,>=latex] (ne1)--(mac7.west);
    \draw[->,>=latex] (ne1)--(mac8.west);
    \draw[->,>=latex] (ne1)--(mac9.west);

    \draw[->,>=latex] (ne2)--(mac10.west);
    \draw[->,>=latex] (ne2)--(mac11.west);
    \draw[->,>=latex] (ne2)--(mac12.west);
    \draw[->,>=latex] (ne2)--(mac13.west);
    \draw[->,>=latex] (ne2)--(mac14.west);

    \draw[->,>=latex] (mac0.east) --(sum0.west)   ;
    \draw[->,>=latex] (mac1.east) --(sum1.west)   ;
    \draw[->,>=latex] (mac2.east) --(sum2.west)   ;
    \draw[->,>=latex] (mac3.east) --(sum3.west)   ;
    \draw[->,>=latex] (mac4.east) --(sum4.west)   ;
    \draw[->,>=latex] (mac5.east) --(sum0.west)   ;
    \draw[->,>=latex] (mac6.east) --(sum1.west)   ;
    \draw[->,>=latex] (mac7.east) --(sum2.west)   ;
    \draw[->,>=latex] (mac8.east) --(sum3.west)   ;
    \draw[->,>=latex] (mac9.east) --(sum4.west)   ;
    \draw[->,>=latex] (mac10.east)--(sum0.west)   ;
    \draw[->,>=latex] (mac11.east)--(sum1.west)   ;
    \draw[->,>=latex] (mac12.east)--(sum2.west)   ;
    \draw[->,>=latex] (mac13.east)--(sum3.west)   ;
    \draw[->,>=latex] (mac14.east)--(sum4.west)   ;

    \draw[->,>=latex] (sum0)--(act0);
    \draw[->,>=latex] (sum1)--(act1);
    \draw[->,>=latex] (sum2)--(act2);
    \draw[->,>=latex] (sum3)--(act3);
    \draw[->,>=latex] (sum4)--(act4);

    \draw[->,>=latex] (act0)--(o0);
    \draw[->,>=latex] (act1)--(o1);
    \draw[->,>=latex] (act2)--(o2);
    \draw[->,>=latex] (act3)--(o3);
    \draw[->,>=latex] (act4)--(o4);
    
\end{tikzpicture} 
    \caption{Data-path of a convolutional layer (bias omitted): The factorization of neighborhood extraction process reduces the memory buffers by a factor of 5 when compared to figure \ref{dhm_layer}}
    \label{layer_post_nef}
\end{figure}
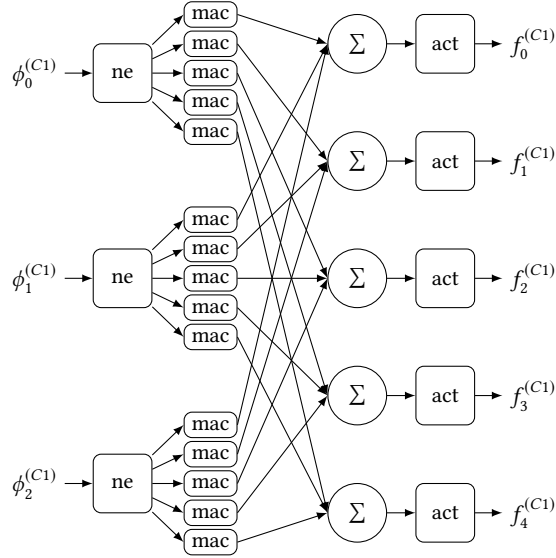

\begin{figure}[!h]
	\centering
	\includegraphics[width=.4\textwidth]{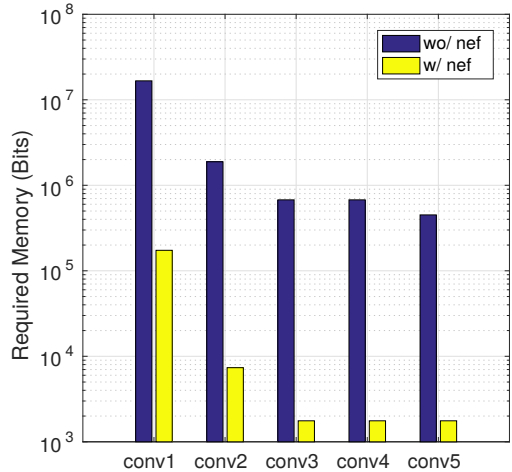}
	\caption{Ratio of memory requirements between architectures w/ and wo/ \ac{nef} for Alexnet convolutional layers: 390\% less memory  is required when factorizing the neighborhood
	extractors}
	\label{nef_alexnet}
\end{figure}

Thus, since multiple neurons in a given layer have same input features to process (only the convolution kernels change), the neighborhood extraction entity can be factorized for each input feature map which divides the memory requirements of each layer by a factor $N^{(l)}$ (cf figure \ref{layer_post_nef}). For instance, while the first layer of the AlexNet CNN (N=96,C=3,K=11) would require $96 \times 3 \times 11 \times 11 = 34KB$ of buffer memory to be processed, a factorization of neighborhood extractors needs $0.3 KB$ which corresponds to 96 times less memory requirements. Full results of \ac{nef} on Alexnet layers are detailed in figure \ref{nef_alexnet}.

%This optimization effect accumulates all along the network's convolutional layers and its efficiency grows with the layer size and network complexity. For instance, in Alexnet convolution layers (with xx neurons) \hl{faire la manip ...} 

\subsection{Constant multiplication}
\subsubsection{Fixed-point computing for CNNs}

Several studies~\cite{suyog15,Gysel2016} have demonstrated that \acp{cnn}, and more generally deep learning applications, usually tolerate approximate computations with short fixed-point arithmetic. Frameworks such as Ristretto~\cite{Gysel2016}, for example, can perform fine-tuning of data representation in order to support fixed-point numerical representations with variable data lengths. In particular, an 8-bit (resp. 2-bit) precision is sufficient to infer the AlexNet~\cite{Krizhevsky2012a} (resp. LeNet~\cite{lecun98}) \acp{cnn} with little to no degradation in classification accuracy. The DHM approach advocated in this work can indeed take advantage of this to significantly reduce the amount of required hardware resources by first inferring the minimal required precision and then \emph{deriving} the size of the hardware resources to exactly match this precision with the adequate bit-width. 

\subsubsection{Multiplications with Logic Elements}
\label{le_arith}
Convolutions require many multiplications. If these multiplications are implemented using hardwired \ac{dsp} blocks within the target \ac{fpga}, this  dramatically limits the complexity of the CNN that can be implemented. For instance, the second layer of the LeNet5 network ($C=6, N=16, K=5$) requires $2400$ multipliers. This number largely exceeds the number of hard-wired multiplier blocks provided by many FPGAs especially by embedded devices. We overcome this problem by systematically forcing the synthesis tool to implement multiplications with logical elements instead of DSP blocks, leading the resulting implementations to rely on AND gates and trees of half-adders~\cite{Altera04}.
% For Xilinx too. only the reference is an Altera application note 

In this case, the logic elements required to implement a convolution increase quadratically with precision. Moreover, due to the large number of multiplications involved in CNNs, the available logic on embedded FPGA devices may not be suffice to support a full complex CNN graph. 
We take advantage of the fact that in the case of CNNs the convolution kernels -- and hence the second operand of the multiplications -- are actually constants and derived from the offline training stage. It is therefore possible to use a specialized version for those multiplier instances. While this approach limits the flexibility of the system -- it requires to re-compile and re-synthesise the VHDL design whenever parameters values are changed --, it delegates to the synthesis tool the task to perform low-level area and performance optimizations. More particularly, multiplications by 0 (\textit{resp} 1) are removed (\textit{resp.} replaced by a simple signal connection) and multiplications by a power of 2 are implemented using shift registers. 

\begin{figure}[h]
    \centering
    \includegraphics[width=.42\textwidth]{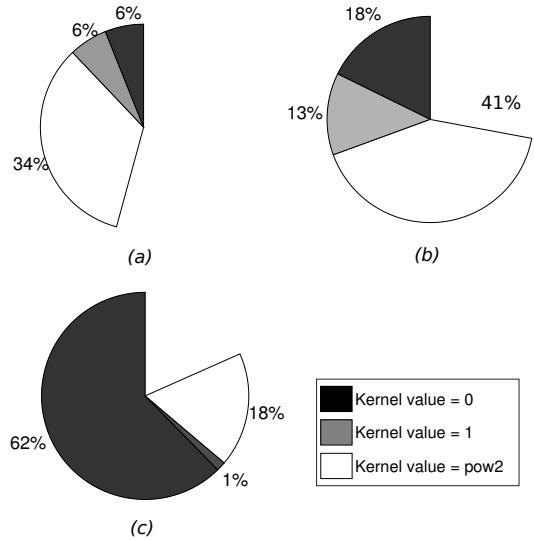}
    \caption{Number of kernels with a null, unitary and power-of-two elements using 8-bits representation. (a)-LeNet5\cite{lecun98}, (b)-bvlc Alexnet\cite{Krizhevsky2012a}, (c)-compressed Alexnet\cite{han2015learning}}
    \label{kernel_stats}
\end{figure}

\begin{figure}[!h]
	\centering
	\includegraphics[width=.5\textwidth]{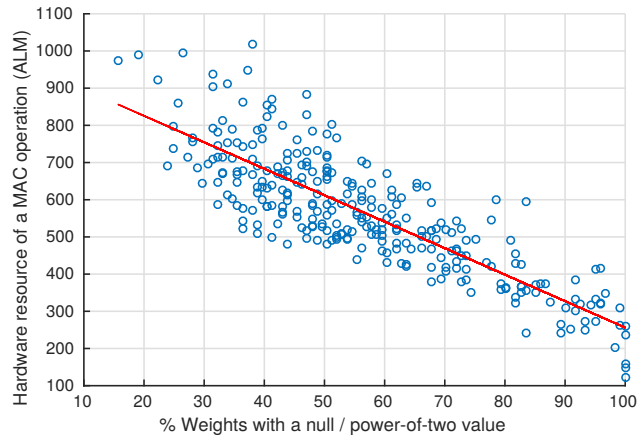}
	\caption{\ac{le} usage of a parallel \ac{mac} ($K=11$): The hardware cost of a \ac{mac} operation decreases with the number of parameters that are equal to a power-of-two. The intercept term (at 100 \%) corresponds to the hardware usage of adder trees}
    \label{alm_vs_nnp}
\end{figure}

Moreover, we find that a large proportion of CNN parameters are, after quantization process, equal to zero, one or a power of two. This is illustrated in figure~\ref{kernel_stats} where 72\% of the AlexNet multiplications (with an 8 bit precision) can be either removed, or replaced with signals or shift registers. Figure~\ref{alm_vs_nnp} shows how the logic elements required to implement a pipelined convolution decrease as the proportion of these "special" kernels increase. 

\begin{table}[!h]
\centering
\label{my-label}
\begin{tabular}{c|c|c|}
\cline{2-3}
                                                         & \multicolumn{2}{c|}{Multiplicand}                                \\ \cline{2-3} 
                                                         & \small{Variable}                & \small{Constant }               \\ \hline 
\multicolumn{1}{|l|}{\multirow{2}{*}{\small{LE Based}}}  & \small{ALM: 380  (0.67 \%)} & \small{ALM : 121  (0.21 \%)}\\ \cline{2-3}   
\multicolumn{1}{|l|}{}                                   & \small{DSP : 0   (0    \%)} & \small{DSP : 0    (0    \%)}\\ \hline
\multicolumn{1}{|l|}{\multirow{2}{*}{\small{DSP Based}}} & \small{ALM : 71  (0.12 \%)} & \small{ALM : 70   (0.12 \%)}\\ \cline{2-3} 
\multicolumn{1}{|l|}{}                                   & \small{DSP : 10  (6.41 \%)} & \small{DSP : 7    (4.48 \%)}\\ \hline
\end{tabular}
\caption{Resource utilization of a random $3 \times 3$ convolution engine on an Altera Cyclone V device with different implementations. }
\end{table}

\section{The \textsc{Haddoc2} utility}

The \textsc{Haddoc2} framework is set of tools built upon the principles and optimization techniques described in the previous section. It is capable of automatically generating a  platform independent hardware description of a \ac{cnn} from a Caffe model~\cite{Jia2014}. First, layer specifications (Layer type, Number of input channels $C$, Number of output features $N$, kernel size $K$) are extracted from the Caffe model and the learned parameters are read, rounded to a fixed-point representation format and written as generic parameters in a configuration file. Second, a top-level VHDL file is created by transforming the dataflow graph described in Caffe. The top-level instantiates a set of generic layers parametrized according to the Caffe model specifications. These layers are described using a small number of basic predefined actors. These actors, written in a structural VHDL, follow the dataflow execution semantics discussed in the latter sections. The output is a platform independent VHDL code that can be implemented on the FPGA device using the adequate synthesis tool. The \textsc{Haddoc2} framework and the library of \ac{cnn} actors supporting the \ac{dhm} approach are open-source and available online\footnote{https://github.com/KamelAbdelouahab/haddoc2}.

\begin{figure}[!h]
    \centering
    \begin{tikzpicture}[scale=0.8, every node/.style={scale=0.8}]

\tikzset{bigB/.style={draw,rectangle,dashed,minimum width=2.5cm,minimum height=2.5cm}};
\tikzset{bbigB/.style={draw,rectangle,minimum width=2cm,minimum height=2.5cm,rounded corners=3pt,fill=gray}};
\tikzset{smallB/.style={draw,rectangle,rounded corners=3pt}};

\node[smallB] (sw_topo) at(0,0.5) {.prototxt};
\node[smallB] (sw_data) at(0,-0.5) {.caffemodel};
\node[smallB] (hw_topo) at(7,0.5) {toplevel.vhd}; 
\node[smallB] (hw_data) at(7,-0.5) {params.vhd}; 

\node[bigB]   (sw)      at(0,0)  {};
\node[bigB]   (hw)      at(7,0)  {};

\node   (sw_wr)      at(0,-2)  {Caffe};
\node   (hw_wr)      at(7,-2)  {Hardware};

\draw[->,>=latex] (sw_topo)--(hw_topo);
\draw[->,>=latex] (sw_data)--(hw_data);

\node[bbigB]  (tool)    at(3.5,0)  {\textbf{\sc{Haddoc2}}};
\end{tikzpicture}
    \caption{Hardware generation: the CNN layer arrangement is described in the top-level files while kernel parameter values and layer specification are written on the configuration file.}
\end{figure}
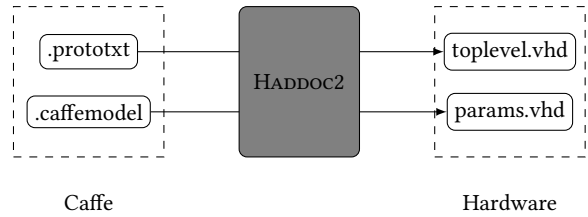

\begin{minipage}[t]{0.45\linewidth}
\begin{lstlisting}[language=json,caption={Caffe description of a $conv$ layer}]
name: "LeNet"

...

layer {
  name: "conv2"
  type: "Convolution"
  bottom: "pool1"
  top: "conv2"
  param {
    lr_mult: 1
  }
  param {
    lr_mult: 2
  }
  convolution_param {
    num_output: 50
    kernel_size: 3
    stride: 1
    weight_filler {
      type: "xavier"
    }
    bias_filler {
      type: "constant"
    }
  }
}

...
\end{lstlisting}
\end{minipage}
\hfill
\begin{minipage}[t]{0.45\linewidth}
\begin{lstlisting}[language=my_vhdl,caption={Generated VHDL code of the layer}]
...
architecture RTL of lenet is

...

conv2: convLayer
  generic map(
    PIXEL_SIZE   => PIXEL_SIZE,
    IMAGE_WIDTH  => CONV2_IMAGE_WIDTH,
    KERNEL_SIZE  => CONV2_KERNEL_SIZE,
    NB_IN_FLOWS  => CONV2_IN_SIZE,
    NB_OUT_FLOWS => CONV2_OUT_SIZE,
    KERNEL_VALUE => CONV2_KERNEL_VALUE,
    KERNEL_NORM  => CONV2_KERNEL_NORM,
    BIAS_VALUE   => CONV2_BIAS_VALUE
  )
  port map(
    clk      => clk,
    reset_n  => reset_n,
    enable   => enable,
    in_data  => pool1_data,
    in_dv    => pool1_dv,
    in_fv    => pool1_fv,
    out_data => conv2_data,
    out_dv   => conv2_dv,
    out_fv   => conv2_fv
  );
    
...
\end{lstlisting}
\end{minipage}

\section{Experimental Results with Haddoc2} \label{sec:res}
As a proof of concept, we have implemented, using the \textsc{Haddoc2} framework,  FPGA-based accelerators for three CNN-based applications, listed in~Table~\ref{res:setup}. The first one is the Caffe version of the LeNet5~\cite{lecun98} \ac{cnn} that requires 20.78 MOPs to process a frame of size 28x28. The second application is the face detector used in~\cite{Farabet2009} which requires 622.08 MOPs to process a 320x240 frame. The last one is introduced in~\cite{Huttunen2016} to perform car type classification and requires 268.28 MOPs to process 96x96 frames. The two first \acp{cnn} have been trained using Caffe while the third model has been directly downloaded as a Caffe pre-trained model. 
Table~\ref{res:setup} gives parameter values for each \ac{cnn} convolutional layer. LeNet5 and CarType \acp{cnn} have 2 convolutional layers while FaceDetect has 3. The corresponding hardware descriptions of each network have been automatically generated using Haddoc2 on an Intel i7-4770 CPU and were synthesised on two FPGA devices using respectively Intel Quartus 16.1 and Xilinx Vivaldo 2016.4. 

\begin{table}[!h]
\centering
\caption{Topology of the convolutional layers  of studied \acp{cnn}.}
\begin{tabular}{l|ccc|ccc|ccc|}
\cline{2-10}
                                    & \multicolumn{3}{|c|}{LeNet5~\cite{lecun98}}  & \multicolumn{3}{|c|}{FaceDetect~\cite{cnp10} } & \multicolumn{3}{|c|}{CarType~\cite{Huttunen2016}}    \\ \hline
\multicolumn{1}{|l|}{Input size}    & \multicolumn{3}{c|}{28 x 28}       & \multicolumn{3}{c|}{320 x 240}        & \multicolumn{3}{c|}{96 x 96 x3}      \\ \hline \hline
\multicolumn{1}{|l|}{Layer parameters}         & $N$         & $C$       & $K$      & $N$         & $C$         & $K$       & $N$         & $C$         & $K$      \\ \hline 
\multicolumn{1}{|l|}{conv1+maxpool} & $20$        & $1$       & $5$      & $6$         & $1$         & $7$       & $32$        & $3$         & $5$      \\ \hline
\multicolumn{1}{|l|}{conv2+maxpool} & $50$        & $20$      & $5$      & $10$        & $6$         & $7$       & $32$        & $32$        & $5$      \\ \hline
\multicolumn{1}{|l|}{conv3}         & $-$         & $-$       & $-$      & $30$        & $10$        & $3$       & $-$         & $-$         & $-$      \\ \hline \hline
\multicolumn{1}{|l|}{Kops/Pixel}    & \multicolumn{3}{c|}{$26.5$}        & \multicolumn{3}{c|}{$6.3$}            & \multicolumn{3}{c|}{$29.1$}          \\ \hline
\end{tabular}
\label{res:setup}
\end{table}

Table~\ref{res:compare} reports post-fitting results of the LeNet-5 accelerator on an embedded Intel Cyclone V 5CGXFC9E7 device using 3 implementation strategies. In the first case, only DSP blocks are used to map the CNN multiplications. The resulting hardware requires $72\times$ the available resource of the device. The second case features an implementation of multiplication based on logic elements and requires $3.8\times$ the available logic. Using tailored multipliers reduces resources by a factor of $8.6\times$, fitting the CNN accelerator onto an Intel Cyclone V device.

\begin{table}[h]
\centering
\caption{Resource utilization by a \ac{dhm} LeNet5 CNN  with different implementations strategies for multipliers.}
\begin{tabular}{c|c|c|c|}
\cline{2-4}
\cline{2-4}
                                        & DSP-based       & LE-based       & LE-based + const. \\ \hline
\multicolumn{1}{|l|}{Logic Usage (ALM)} & NA              & 433500 (381\%) & 50452 (44\%)     \\ \hline
\multicolumn{1}{|l|}{DSP Block usage}   & 24480 (7159 \%) & 0 (0\%)        & 0 (0\%)          \\ \hline                    
\end{tabular}
\label{res:compare}
\end{table}

Table~\ref{res:fit} details post fitting results on two embedded \ac{fpga} platforms: the Intel Cyclone V 5CGXFC9E7 and the Xilinx Kintex7 XC7Z045FBG. To the best of our knowledge, these numbers are the first to demonstrate the applicability of a DHM-based approach for the implementation of \acp{cnn} on embedded FPGAs. The three hardware accelerators fit onto the embedded devices with no off-chip memory requirement. The memory footprint shown in post fitting reports corresponds to line buffers used by the dataflow-based convolution engine and both synthesis tools instantiate LUT-based memory blocks to implement these buffers. As expected when using \ac{dhm}, the logic utilization in the \ac{fpga} grows with the the topology of the \ac{cnn}. However, in all the studied cases, the resources are sufficient to support direct hardware mapping. Finally, the same table reports timing analysis results of the three generated hardware accelerators. With a peak frequency of 62.3 MHz for the CarType \ac{cnn}, \ac{dhm} grants a maximum computation throughput of 1813 GOPs/s. For the face detection neural network, the presence of a third convolutional layer in the pipeline drops the maximum frequency to 56.7 MHz (i.e 357 GOPs/s) in the Cyclone device, which corresponds to 164 classifications/sec on 512x512 images with a 3-multiscale pyramid.

%\footnotetext[2]{HDL was generated on an Intel i7-4770 CPU}
%\footnotetext[3]{Synthesised using Quartus Prime 15.1} 

\begin{table}[h]
\centering
\caption{Resource Utilization of the Haddoc2-generated convolutional layers of studied CNNs  with 5-bit representation on: a- an Intel Cyclone V FPGA,  b- a Xilinx Kintex 7 FPGA.}
\begin{tabular}{ll|c|c|c|}
\cline{3-5}
                                        &                                 & LeNet5~\cite{lecun98} & FaceDetect~\cite{cnp10} & CarType~\cite{Huttunen2016}  \\ \hline                                  
\multicolumn{1}{|l|}{\multirow{5}{*}{a}}& Logic Elements (ALMs)           & 50452 (44\%)         & 6158  (5\%)                  &  48243 (42\%)                  \\ \cline{2-5} 
\multicolumn{1}{|l|}{}                  & DSP Blocks\footnotemark[1]      & 0     (0 \%)         & 0   (0\%)                    &  0     (0\%)                   \\ \cline{2-5} 
\multicolumn{1}{|l|}{}                  & Block Memory Bits               & 2752  (1\%)          & 41408 (1\%)                  &  28320 (1\%)                   \\ \cline{2-5}
\multicolumn{1}{|l|}{}                  & Frequency                       & 69.14  MHz           & 56.7  MHz                    &  66.0  MHz                     \\ \cline{2-5}
\multicolumn{1}{|l|}{}                  & Processing capabilities         & 1832 GOPs/s          & 357   GOPs/s                 &  1920  GOPs/s                  \\ \hline \hline
\multicolumn{1}{|l|}{\multirow{5}{*}{b}}& Slices                          & 48114 (88\%)         & 6221  (11\%)                 &  49082 (89\%)                  \\ \cline{2-5} 
\multicolumn{1}{|l|}{}                  & DSP Blocks\footnotemark[1]      & 0     (0\%)          & 0     (0\%)                  &  0     (0\%)                   \\ \cline{2-5} 
\multicolumn{1}{|l|}{}                  & LUTs as Memory                  & 420   (1\%)          & 1458  (2\%)                  &  1154  (1\%)                   \\ \cline{2-5}
\multicolumn{1}{|l|}{}                  & Frequency                       & 62.13 MHz            & 44.41 MHz                    &  62.3  MHz                     \\ \cline{2-5}
\multicolumn{1}{|l|}{}                  & Processing capabilities         & 1646  GOPs/s         & 279   GOPs/s                 &  1813 GOPs/s                   \\ \hline
\end{tabular}
\label{res:fit}
\end{table}
\normalsize
\section{Related work}
Several studies leverage on FPGA computational power and hardware flexibility to implement the feed-forward propagation of \acp{cnn}. A non exhaustive review of these can be found in~\cite{Lacey2016}. In most of approaches, acceleration of CNN-based applications is provided by mapping a limited subset of processing elements onto the target device. This is the case for example in~\cite{microsoft15} where authors describe an accelerator for the AlexNet \ac{cnn}~\cite{Krizhevsky2012a} implemented on a large Stratix V FPGA which, to the best of our knowledge, outperforms most state-of-the-art implementations in terms of computational and outperformed most of state-of-the-art implementations such \cite{chakra10,peeman13,cnp10}.  Most of these designs are \ac{fpga} based accelerators for convolution with a relatively similar architecture of parallel processing elements associated with embedded hardcore processors running a software layer. Other approaches like \cite{zhang15} relies on analytical design scheme using the roofline model and loop tiling to propose an inference engine where the attainable computation roof of the \ac{fpga} is reached. This loop tilling optimization is performed on a C code then implemented in floating point on a Virtex 7 485T using Vivaldo HLS Tool.\\

% \subsection{Dataflow Oriented Implementation}
As it has been seen in the latter sections, feed forward propagation is an algorithm that intrinsically suits to dataflow processing. Thus, dedicated stream processors for \acp{cnn} have been proposed. The most notable contribution was neuFlow \cite{neuflow}: A runtime reconfigurable processor for real-time image classification. In this work, Farabet and al. introduced a grid of processing tiles that were configured on runtime to build a dataflow graph for \ac{cnn} applications. It was associated to "luaFlow": a dataflow compiler that transforms a high-level flow-graph representation of an algorithm into machine code for neuFlow. Such architecture was implemented on a Virtex 6 VLX240T and provided a 12 fps categorization for 512x375 images. Thus, NeuFlow transformed a \ac{cnn} graph into a set of dataflow instructions, where each instruction is described as an hardware configuration of 2D-processing elements called \emph{Processing tiles (PTs)}. Execution of the graph is carried out by sequencing the instructions on the target FPGA. This approach requires an external memory to store intermediate results, which in turn, even with the help of a DMA, limits the final speedup. The study in~\cite{Venieris2016} features a partitioning of the \ac{cnn} graph with one bitstream per subgraph in a way that only on-chip memory is needed to store intermediate results. This however requires the reconfiguration of the FPGA whenever data has to enter a different subgraph, which adds a substantial reconfiguration time overhead.\\

By contrast, the DHM approach and \textsc{Haddoc2} tool introduced in the present work performs all processing \emph{on the fly} and does not require an external memory to store intermediate results. Throughput is therefore not limited by off-chip memory bandwidth. Previous works in \cite{abdelouahab16} describe a first version of Haddoc that relied on the Caph \cite{CAPH} , a \ac{hls} tool to provide dataflow-based hardware accelerators for \acp{cnn} on \acp{fpga}. While this implementation operated at very high frame-rates (800 classifications/sec on 256 $\times$ 256 images), the  over-head that comes with the \ac{hls} heavily restrained the size of \acp{cnn} to be implemented which motivated us to bypass the Caph HLS layer by hand-crafting RTL IP cores that respects the dataflow execution model and supports the detailed DHM concepts.

\bibliographystyle{unsrt}
\bibliography{biblio}

\end{document}